# Antiferromagnetic ordering in van der Waals two-dimensional magnetic material MnPS$_3$ probed by Raman spectroscopy

*Kangwon Kim,[†] Soo Yeon Lim,[†] Jungcheol Kim,[†] Jae-Ung Lee,[†,#] Sungmin Lee,[‡,§] Pilkwang Kim,[§,∥] Kisoo Park,[‡,§] Suhan Son,[‡,§] Cheol-Hwan Park,[§,∥,*] Je-Geun Park,[‡,§,*] and Hyeonsik Cheong[†,*]*

[†]Department of Physics, Sogang University, Seoul 04107, Korea

[#]Department of Physics, Ajou University, Suwon 16499, Korea

[‡]Center for Correlated Electron Systems, Institute for Basic Science, Seoul 08826, Korea

[§]Department of Physics and Astronomy, Seoul National University, Seoul 08826, Korea

[∥]Center for Theoretical Physics, Seoul National University, Seoul 08826, Korea

**\*E-mail:** cheolhwan@snu.ac.kr.

**\*E-mail:** jgpark10@snu.ac.kr.

\***E-mail:** hcheong@sogang.ac.kr

**Abstract**

Magnetic ordering in the two-dimensional limit has been one of the most important issues in condensed matter physics for the past several decades. The recent discovery of new magnetic van der Waals materials heralds a much-needed easy route for the studies of two-dimensional magnetism: the thickness dependence of the magnetic ordering has been examined by using Ising- and XXZ-type magnetic van der Waals materials. Here, we investigated the magnetic ordering of MnPS$_3$, a two-dimensional antiferromagnetic material of Heisenberg-type, by Raman spectroscopy from bulk all the way down to bilayer. The phonon modes that involve the vibrations of Mn ions exhibit characteristic changes as temperature gets lowered through the Néel temperature. In bulk MnPS$_3$, the Raman peak at ~155 cm$^{-1}$ becomes considerably broadened near the Néel temperature and upon further cooling is subsequently red-shifted. The measured peak positions and polarization dependences of the Raman spectra are in excellent agreement with our first-principles calculations. In few-layer MnPS$_3$, the peak at ~155 cm$^{-1}$ exhibits the characteristic red-shift at low temperatures down to the bilayer, indicating that the magnetic ordering is surprisingly stable at such a thin limit. Our work sheds light on the hitherto unexplored magnetic ordering in the Heisenberg-type antiferromagnetic systems in the atomic-layer limit.

## 1. Introduction

Two-dimensional (2D) van der Waals magnetic materials are attracting intense interest not only for their technological importance but also because they can address the fundamental question of magnetism in low-dimensional systems [1]. Typically, strong fluctuations can easily destroy magnetic ordering in low-dimensional systems. For example, no magnetic ordering is possible in one dimension [2]. 2D systems are much more interesting because the long-range order depends on both the symmetry of the order parameter and the type of spin-spin interactions, which compete with intrinsic fluctuations of either quantum and/or thermal nature.

The generic magnetic Hamiltonian for such systems can be written as [3]

$$H = -J_{XY}\sum_{j\delta}(S_j^x S_{j+\delta}^x + S_j^y S_{j+\delta}^y) - J_I \sum_{j\delta} S_j^z S_{j+\delta}^z, \qquad (1)$$

where $J_{XY}$ and $J_I$ are spin-exchange energies on the basal plane and along the *c*-axis, respectively; $S_j^\alpha$ is the $\alpha$ ($\alpha = x, y$, or $z$) component of total spin; and *j* and *δ* run through all lattice sites and all nearest-neighbors, respectively. All three fundamental models can be realized with the generic Hamiltonian: $J_{XY} = 0$ for the Ising model, $J_I = 0$ for the XY model, and $J_{XY} = J_I$ for the Heisenberg model. According to the Mermin-Wagner theorem [4], no magnetic ordering is possible at any nonzero temperature in one- or two-dimensional isotropic Heisenberg models. On the other hand, 2D Ising systems can have magnetic ordering at finite temperatures according to Onsager [5].

Transition metal phosphorus trisulfides (TMPS$_3$) belong to a class of 2D van der Waals magnetic materials that can be exfoliated to atomically thin layers [6,7]. For transition metal elements like Fe, Ni, and Mn, the materials share the same crystal structures but the magnetic phase at low temperatures vary depending on the magnetic elements: Ising (Fe), XXZ (Ni), and Heisenberg (Mn) antiferromagnets, respectively [8]. It therefore offers a unique opportunity to investigate the magnetic ordering behavior in the 2D limit for different types of exchange interactions.

Unfortunately, conventional tools such as neutron scattering is not suitable for atomically thin layers of these compounds due to a very small sample volume [1,9]. Moreover, because antiferromagnetic phases do not have net magnetization, a direct measurement of antiferromagnetic ordering using a tool such as magneto-optical Kerr effect (MOKE), which is otherwise useful in the case of atomically-thin ferromagnetic materials, is not possible, either [10,11]. Raman spectroscopy, on the other hand, has proven to be a powerful tool to investigate magnetic ordering in atomically thin 2D magnetic materials [12–16]. A simple reason is that some of the phonon modes can be coupled with magnetic ordering to exhibit characteristic changes across the critical temperature, or other Raman features due to magnetic structure such as magnons can be correlated with magnetic ordering. In the lack of more direct measurement techniques such as MOKE for antiferromagnetic materials, Raman spectroscopy has therefore become the most important tool in probing the antiferromagnetic ordering in the 2D limit. For example, Raman spectroscopy has been recently utilized to demonstrate that indeed antiferromagnetic ordering is sustained down to the 2D limit of monolayer $FePS_3$ [13,14]. In the case of XXZ-type $NiPS_3$, many Raman spectroscopic features indicate that the magnetic ordering occurs down to the bilayer (2L) with a slight decrease of the Néel temperature as the thickness is decreased, but is strongly suppressed in the monolayer [16]. Here, we report on the Raman spectroscopic analysis of the Heisenberg-type antiferromagnet $MnPS_3$ using Raman spectroscopy. We discovered a unique feature of the Raman spectrum that correlates well with the antiferromagnetic ordering. We further show that such magnetic ordering is surprisingly observed down to bilayer $MnPS_3$

## 2. Method

### 2.1 Crystal growth and characterization.

Single crystals of MnPS$_3$ were grown by a self-flux chemical vapor transport method in quartz tube ampoule evacuated to high vacuum. Manganese powder (99.95%, Alfa Aesar), red phosphorous (99.99%, Sigma-Aldrich) and sulfur flakes (99.99%, Sigma-Aldrich) were mixed in the stoichiometric ratio with 5 wt% of extra sulfur within Ar atmosphere (< 1 ppm of moisture and oxygen). The mixture was loaded in a quartz tube ampoule (10 mm of inner diameter and 150 mm in length) and sealed at the pressure of ~$10^{-2}$ Torr. The sealed tube was placed in two-zone furnace and heated to 780°C (containing powder mixture) / 720°C for 7 days. The quartz tube was cooled to room temperature during 2 days. Single crystals are found as green and transparent plates with the typical size of 10 mm×10 mm×100 μm. After additional 1 day annealing under Ar atmosphere to remove extra sulfur, we verify the sample stoichiometry with energy dispersive X-ray spectroscopy (EDX). The measurements of magnetic properties were performed using a SQUID magnetometer (Quantum Design, MPMS3) as shown in figure 2(e).

### 2.2 Raman measurements.

The temperature dependent Raman spectra of a bulk MnPS$_3$ crystal was measured in a macro-Raman system by using a closed-cycle He cryostat. The 488-nm (2.54 eV) line of an Ar$^+$ laser was used as the excitation source for all the measurements except for the excitation energy-dependent Raman measurements (Figure S9) and the temperature-dependent Raman measurements with the 2.41-eV excitation energy (Figure S12). The excitation laser was focused by a spherical lens ($f$=75 mm) to a spot of size ~50 μm with a power of 1 mW. For measuring polarized Raman spectra of bulk and Raman spectra of few-layer MnPS$_3$ at low and room temperatures, a micro-Raman system with backscattering geometry were used with samples prepared on Si substrates with a layer of 90-nm $SiO_2$ by mechanical exfoliation. The atomically thin samples are relatively stable in air but can be degraded when the samples are exposed in ambient condition for more than a week. After exfoliation, the samples were kept in a vacuum desiccator to prevent possible degradations. All the micro-Raman measurements were performed in vacuum using an optical cryostat (Oxford Micorostat He2) at temperatures of 10 and 290 K. The laser beam was focused onto the sample by a 40× microscope objective lens (0.6 N.A.), and the scattered light was collected and collimated by the same objective. The laser power was kept below 100 µW in order to avoid local heating. The scattered signal was dispersed by a Jobin-Yvon Horiba iHR550 spectrometer (2400 grooves/mm) and detected with a liquid-nitrogen-cooled back-illuminated charge-coupled-device (CCD) detector. An achromatic half-wave plate was used to rotate the polarization of the linearly polarized laser beam to the desired direction. The analyzer angle was set such that photons with polarization parallel to the incident polarization pass through. Another achromatic half-wave plate was placed in front of the spectrometer to keep the polarization direction of the signal entering the spectrometer constant with respect to the groove direction of the grating. The crystal axes of the samples were determined by comparing X-ray diffraction measurements with polarized Raman measurements. In particular, the polarization dependences of $P_3$ and $P_5$ were used to find the *a* and *b* axes in the plane (see supplementary information figure S1). For few-layer samples, the background signal from the substrate was subtracted from the Raman spectra. For measurements where the polarization dependence is not critical, we removed the analyzer in order to maximize the signal. Nevertheless, because the excitation laser is linearly polarized and the spectrometer efficiency has a polarization dependence, the polarization condition can be described as 'partially parallel-polarized.'

## 2.3 Calculation details.

We calculated the vibrational modes of MnPS$_3$ by diagonalizing the Hessian matrix obtained from analytic derivatives of the total energy (obtained from first principles using density functional theory) with respect to ionic displacements [17,18]. The exchange-correlation energy was calculated by using the Perdew−Burke−Ernzerhof functional [19] with 12.5% of Hartree-Fock exchange energy. The Brillouin zone was sampled with 8 × 8 × 6 Monkhost-Pack *K*-point mesh[20] and the polarized triple-zeta (pob-TZVP) basis set [21] was employed with all-electron core potentials. We optimized both the lattice parameters and the nuclear coordinates. Raman intensities were calculated by using the coupled-perturbed Hartree-Fock/Kohn-Sham (CPHF) approach [22,23]. All our calculations were done with CRYSTAL 17 package [24,25].

*Spin wave calculation* We calculated the one-magnon spectra of MnPS$_3$ using a Heisenberg-type spin Hamiltonian with a single-ion easy-axis anisotropy refined from the previous inelastic neutron scattering study on MnPS$_3$ [26]:

$$H = \sum_{i,j} J_{ij} \mathbf{S}_i \cdot \mathbf{S}_j + \Delta \sum_i (S_i^z)^2 \quad (2)$$

The isotropic exchange interactions, $J_{ij}$, up to the 3$^{rd}$ in-plane nearest neighbor are labelled $J_1 - J_3$. The spin Hamiltonian was diagonalized using the SpinW package [27] for the antiferromagnetic ground state. After obtaining one-magnon energies at randomly chosen $10^6$ sample **q** points throughout the full Brillouin zone, we calculated two-magnon DOS satisfying the following sum rule:

$$D(\mathbf{k}, \omega) = \pi \sum_{\mathbf{q},m,n} \delta\left(\hbar\omega - \hbar\omega_m(\mathbf{q}) - \hbar\omega_n(\mathbf{k} - \mathbf{q})\right) \quad (3)$$

where $\omega_m(\boldsymbol{q})$ is the energy of the *m*-th magnon band.

## 3. Results and Discussion

Bulk TMPS$_3$ crystals have a monoclinic structure with the point group of $C_{2h}$ (2/m) [28–30]. As shown in figure 1(a), the transition metal (TM) atoms form a honeycomb lattice and are surrounded by six S atoms [28–30]. These S atoms themselves are connected to two P atoms above and below the TM plane like a dumbbell. Bulk MnPS$_3$ exhibits Heisenberg-type antiferromagnetic ordering with the Néel temperature of 78 K. Figure 1(b) shows the representative polarized Raman spectra of bulk MnPS$_3$ at 290 and 10 K measured in parallel- and cross-polarization configurations. The room-temperature spectrum is similar to the previously reported (unpolarized) Raman spectrum [29]. The high frequency modes, P$_3$, P$_4$, P$_5$ and P$_6$, are mostly due to the molecular-like vibrations from the $(P_2S_6)^{4-}$ bipyramid structures and are similar to the corresponding modes observed in FePS$_3$ or NiPS$_3$. On the other hand, the low-frequency peaks, P$_1$ and P$_2$, are from vibrations involving the transition metal Mn atoms [29,31] and reflect different atomic masses and magnetic structures from those of FePS$_3$ or NiPS$_3$ (see supplementary information figure S3). The calculated results of the corresponding vibrational modes for MnPS$_3$ can be found in figure 1(c) and supplementary information (table S1, figures S4 and S5). The discrepancy in the frequencies of the Raman spectra obtained from calculation and experiment is as small as ~ 5 cm$^{-1}$ in the case of the most prominent mode (P$_6$) and less than 15 cm$^{-1}$ (P$_4$). This discrepancy is small from the standard of state-of-the-art first-principle calculations on similar transition-metal compounds [13,32–35]. Moreover, theory and experiment show remarkable agreement in the Raman intensities and their light-polarization dependences. Based on this agreement, we can assign the corresponding lattice vibrational mode to each peak in the measured Raman spectra. Unlike the case of FePS$_3$ or NiPS$_3$, where the Raman spectra change dramatically as the temperature is lowered through the Néel temperature, the changes in the Raman spectra are rather subtle for MnPS$_3$. For example, the intensity of P$_1$ decreases substantially and P$_2$ is red-shifted in the spectra taken at 10 K with respect to the ones measured at room temperature. As we will see below, the shift of P$_2$ correlates well with the antiferromagnetic ordering whereas it is more difficult to establish an immediately recognizable correlation between the intensity of P$_1$ and the magnetic ordering because of the low intensity of this peak. The small splitting of P$_3$ and P$_5$ might indicate a deviation from the three-fold rotational symmetry, which can be attributable to the interlayer interaction. The inset of figure 1(b) shows that the P$_3$ splitting changes little between 290 and 10 K. Figures 1(d) and 1(e) show that the position of P$_5$ varies slightly with the incident polarization, indicating that this peak is a superposition of two peaks with orthogonal polarizations dependences (see supplementary information figure S6 for full polarization dependence of P$_5$), which is similar to what is observed in the corresponding Raman modes of FePS$_3$ [13] and NiPS$_3$ [16]. Due to the shifted stacking of the layers as shown in figure 1(a), multilayer MnPS$_3$ has a monoclinic structure which has in-plane anisotropy. This anisotropy leads to the splitting of P$_3$ and P$_5$ which are isotropic E$_g$ modes in a monolayer. The Raman tensor analyses for the modes are included in supplementary information (note S1 and figures S1 and S2).

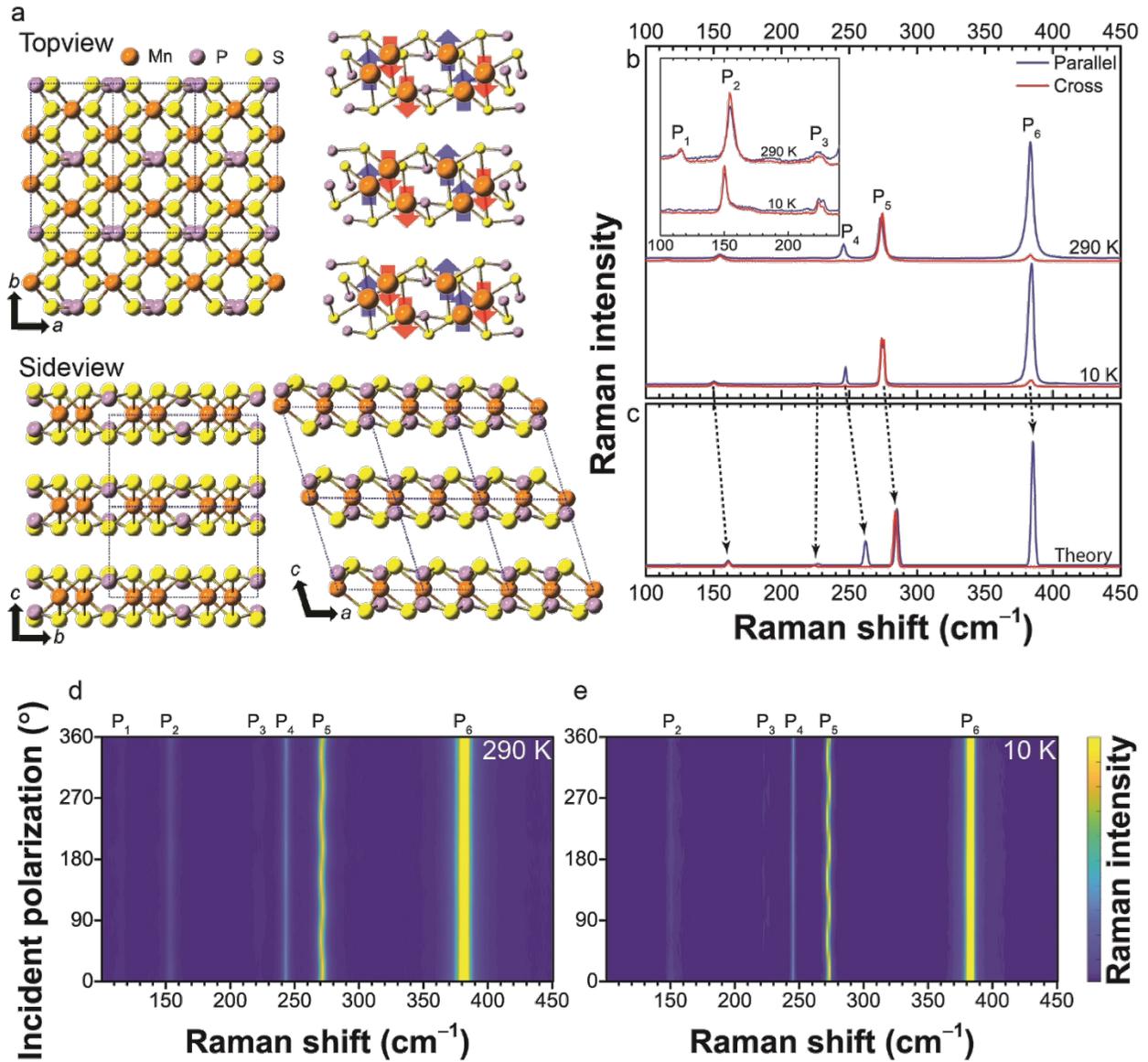

**Figure 1.** (a) Crystal and magnetic structures of MnPS$_3$. (b) Polarized Raman spectra of bulk MnPS$_3$ at 10 and 290 K in parallel and cross polarization configurations. (c) Calculated polarized Raman spectra of bulk MnPS$_3$ in parallel and cross polarization configurations. (d,e) Incident polarization dependence of Raman spectra at (d) 290 and (e) 10 K.

The correlation between the Raman spectrum and the antiferromagnetic phase transition is further investigated by measuring the Raman spectrum of a bulk crystal as a function of temperature as the sample is cooled down through the Néel temperature. Figure 2(a) shows the Raman spectra in the vicinity of $P_2$ measured as a function of temperature in 10 K steps (see supplementary information figure S7 for temperature dependence of the full spectrum). As the temperature is lowered from room temperature, this peak slightly blue-shifts due to the usual volume contraction effect and becomes somewhat sharper owing to the suppression of anharmonic effects at lower temperatures. As the temperature is further lowered through the Néel temperature at 78 K, several additional changes are observable: the peak becomes significantly broader and the peak height is decreased. For temperatures below the Néel temperature, the peak becomes even sharper and, at the same time, is shifted toward lower frequencies. Since the other peaks on higher frequencies ($P_3$ to $P_6$) do not show any abrupt changes in this temperature range (see supplementary information figure S8), these changes are not likely to be caused by a structural phase transition. These changes are summarized in figures 2(b)-(d) and compared with the susceptibility data in figure 2(e). It is immediately clear that there is a very good correlation between our data: the changes in the Raman spectrum, i.e., the shift of the peak frequency [figure 2(b)] and the increase of the full width at half maximum [figure 2(c)], and the antiferromagnetic ordering measured from the magnetic susceptibility. This is reasonable because the $P_2$ mode has a significant weight from the Mn vibrations (see supplementary information figure S4). There is an offset of ~10 K between the Raman and susceptibility data probably because the actual temperature at the measurement spot in Raman experiments is slightly higher than at the thermometer in our cold-finger type closed-cycle cryostat due to temperature gradient inside the cryostat and local heating by the laser. Figure 2(d) shows the spectral weight in the range of 140-180 cm$^{-1}$, which increase more or less monotonically as the temperature is decreased, with some variations near the Néel temperature. The lack of any dramatic change in the spectral weight as the temperature is varied through the Néel temperature indicates that no new scattering channel opens as a result of the magnetic transition.

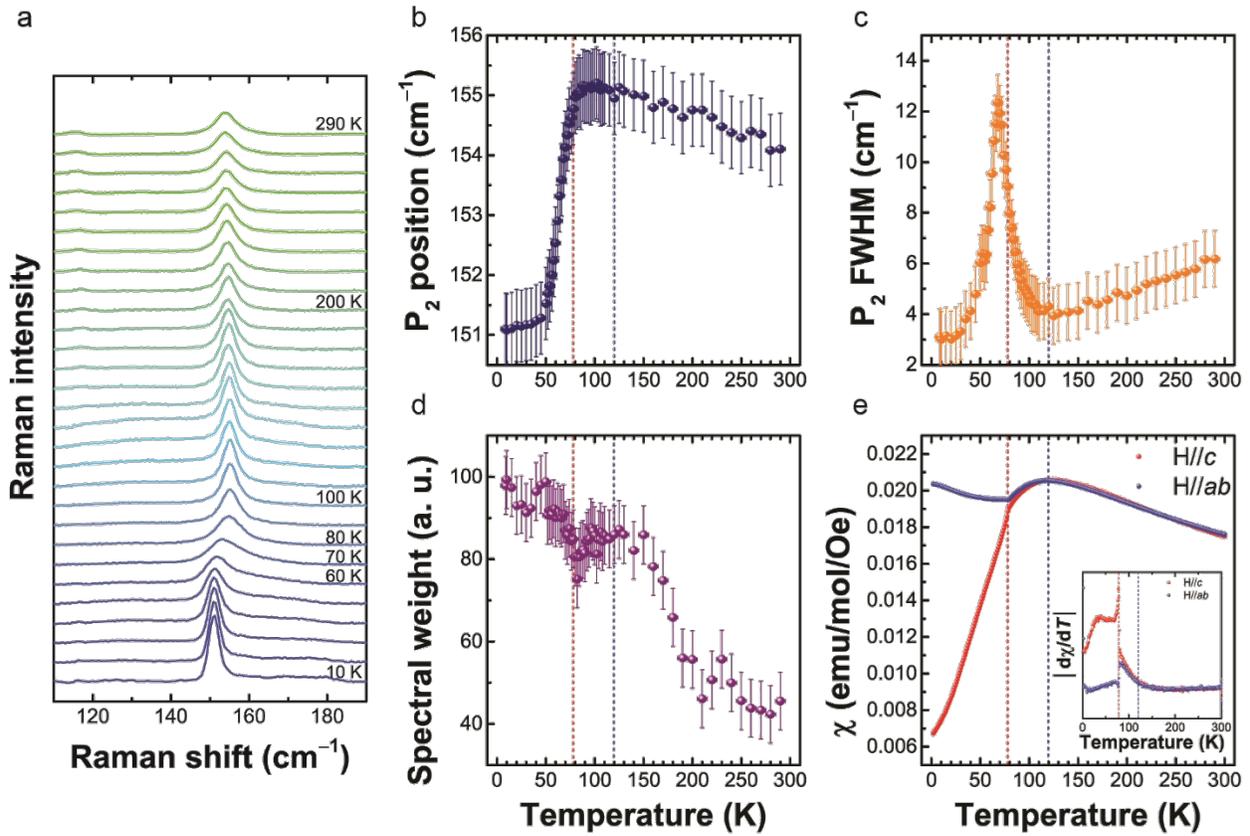

**Figure 2.** (a) Temperature dependence of $P_2$ for bulk $MnPS_3$, measured in the partially parallel-polarized configuration. Temperature dependence of (b) peak position, (c) FWHM (full width at half maximum) of $P_2$, and (d) spectral weight of the spectra in range of 140-180 cm$^{-1}$. (e) Magnetic susceptibility of bulk $MnPS_3$. The inset is the first derivative of the magnetic susceptibility. Red and blue dashed lines indicate the Néel temperature and the temperature at which the susceptibility is maximum, respectively.

A full understanding of the intriguing behavior of $P_2$ across the Néel temperature requires further analysis. One may speculate that it is due to a coupling with the two-magnon scattering that becomes significant below the Néel temperature. The position of this feature approximately coincides with the two-magnon frequency: inelastic neutron scattering measurements found the zone-boundary magnon energy of ~11 meV (~89 cm$^{-1}$) [26]. By using the parameters obtained from ref. [26], we calculated the spin-wave dispersion and the two-magnon density of states (DOS) of MnPS$_3$ (see supplementary information figure S9). The calculated two-magnon DOS have a strong peak near 180 cm$^{-1}$. In the Raman spectra measured with different excitation energies, indeed there is a somewhat broad peak at ~178 cm$^{-1}$ (see supplementary information figure S10). However, the intensity of this feature at the frequency of $P_2$ (~155 cm$^{-1}$) is very weak, which implies that the coupling with this broad feature, presumably due to the two-magnon scattering, is not the major cause of the dramatic changes in the position and line shape of $P_2$ near the Néel temperature. Furthermore, the smooth, monotonic temperature dependence of the spectral weight of $P_2$ [Figure 2(d)] around the Néel temperature also indicates that interaction with two-magnon scattering is less likely the main cause of the red-shift because the contribution of two magnon scattering opens an additional Raman channel that usually results in an abrupt increase in the spectral weight of the Raman spectrum through the Néel temperature [36,37]. On the other hand, MnPS$_3$ is known to have an intermediate phase above the Néel temperature, which is characterized by short-range spin-spin correlations below 120 K, where the susceptibility has a maximum [8,38]. Figure 2(c) shows that the linewidth increases sharply below ~120 K and displays a maximum at or near the Néel temperature. The peak frequency, on the other hand, starts to red-shift below ~80 K. In general, a phase transition is accompanied by large fluctuations near the critical temperature.

Such fluctuations should disrupt the coherence of the lattice vibration modes, shortening the phonon lifetime. The observed temperature dependence of the linewidth can be explained as a combination of the overall monotonic decrease due to suppression of anharmonic effects at low temperatures and the sharp increase of the linewidth due to fluctuations near the phase transition. A similar trend of increase was observed in the linewidth of the corresponding Raman mode of FePS$_3$ near the Néel temperature, although less dramatically (see supplementary information figure S11). This broadening of the Raman peak and its subsequent redshift as the temperature is lowered below the Néel temperature is attributed to spin-phonon coupling, namely, the change of the phonon energy due to magnetic ordering.

We should also note that the line shape of P$_2$ becomes significantly asymmetric near the Néel temperature, reminiscent of a Fano resonance. However, since the two-magnon continuum has little overlap with this phonon energy, it is not likely that the asymmetry is a result of a Fano-like resonance with the magnon band. We suspect that the asymmetry is related to the disruption of the phonon coherence near the phase transition due to the large fluctuations. Another change in the Raman spectrum is the apparent disappearance of P$_1$ at low temperature as shown in the inset of figure 1(b). This peak is better resolved in the spectra measured with the 2.41-eV excitation energy and seems to have similar temperature dependences of the position and FWHM as P$_2$ (see supplementary information note S2, figures S12, and S13). We also find that P$_6$ becomes slightly asymmetric at low temperatures [figure 3(c)]. The other peaks (P$_3$ to P$_6$) do not show any abrupt changes in the line shape or the intensity near the Néel temperature except that the FWHM of P$_6$ increases at low temperatures due to the asymmetry of the peak [see supplementary information figure S8(k)]. Since this mode is due to a breathing-like vibration of the (P$_2$S$_6$)$^{4-}$ bipyramid structures (A$_g$), a magnetic-ordering-induced splitting is not possible. At the same time, polarized

Raman scattering measurements show that the asymmetric line shape does not vary with polarization (see supplementary information figure S14). The origin of this asymmetric line shape at low temperature needs further investigation.

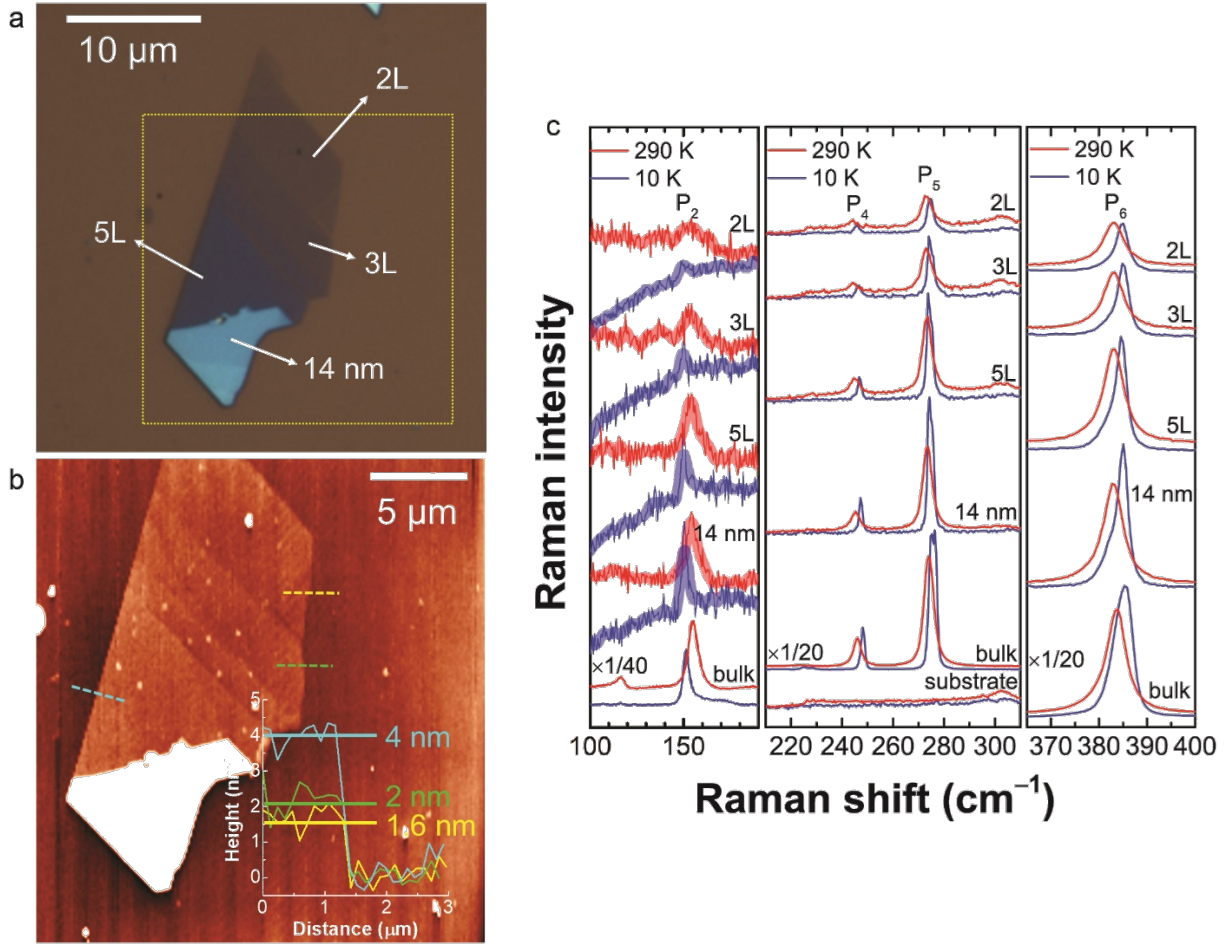

**Figure 3.** (a) Optical contrast and (b) atomic force microscope images of few-layer MnPS$_3$ on SiO$_2$/Si substrate. (c) Raman spectra of few-layer and bulk MnPS$_3$ at 290 and 10 K, measured in the partially parallel-polarized configuration.

In order to investigate the dependence of the magnetic ordering on the number of layers, we measured the Raman spectra of several exfoliated samples at both room temperature and 10 K. Figures 3(a) and (b) show the optical and atomic force microscope images of the sample measured.

Bilayer, trilayer (3L), 5-layer (5L), and a thick 14-nm regions are readily identified in the figures. We also compare the Raman spectra taken at 290 and 10 K for different number of layers in figure 3(c). The spectra from a very thick (~1 μm), bulk-like exfoliated sample are also plotted for comparison. Because of the very small intensity of $P_2$, it is rather difficult to pinpoint the phase transition temperature precisely unlike the case of the bulk sample. However, the comparison shows convincingly enough that whereas other peaks blue-shift in going from 290 to 10 K, $P_2$ shows a noticeable red-shift for all thicknesses, implying that the magnetic ordering is surprisingly sustained down to 2L. For monolayer, the $P_2$ signal could not be resolved due to extremely weak intensity (see supplementary information note S3 and figure S15). For comparison, the magnetic ordering is found down to the monolayer limit for Ising-type $FePS_3$ [13], whereas in XXZ-type $NiPS_3$ it is only stable down to 2L before being completely suppressed in monolayer [16] In Heisenberg-type $MnPS_3$, the magnetic ordering is supposedly more susceptible to fluctuations due to its isotropic nature and thus expected to be more fragile in the 2D limit than in XXZ-type $NiPS_3$. Our experimental results indicate that the Heisenberg-type antiferromagnet $MnPS_3$ can still host magnetic ordering down to two layers.

4. **Conclusions**

We investigated the thickness dependence of the magnetic phase transition in Heisenberg-type two-dimensional antiferromagnetic material $MnPS_3$ by Raman spectroscopy. The phonon modes that involve the vibrations of the Mn ions exhibit characteristic changes as the temperature is lowered through the Néel temperature. In bulk $MnPS_3$, the Raman peak at ~155 cm$^{-1}$ becomes considerably broader near the Néel temperature and is red-shifted below the Néel temperature. Another weak peak at ~117 cm$^{-1}$ also shows similar changes in the line shape near the Néel temperature, although the changes are less dramatic. Using these changes in the phonon modes induced by the magnetic ordering, we found that the magnetic ordering is surprisingly stable down to the bilayer MnPS$_3$.

**Acknowledgements**

This work was supported by the National Research Foundation (NRF) grants funded by the Korean government (MSIT) (NRF-2019R1A2C3009189 and No.2017R1A5A1014862, SRC program: vdWMRC center, No. 2016R1A1A1A05919979), by a grant (2013M3A6A5073173) from the Center for Advanced Soft Electronics under the Global Frontier Research Program of MSIT, and by the Creative-Pioneering Research Program through Seoul National University. Computational resources have been provided by KISTI Supercomputing Center (KSC-2018-C2-0002). The work at IBS CCES was supported by the Institute of Basic Science (IBS) in Korea (Grants No. IBS-R009-G1).

**Author contributions**

J.-G.P. and H.C. conceived the experiments. S.L. and S.S. grew and characterized bulk MnPS$_3$ crystals. K.K., S.Y.L., J.K., and J.-U.L. carried out Raman measurements. P.K. carried out first-principles calculations on the lattice vibrations and Raman spectra. K.P. calculated the spin waves. All authors discussed the data and wrote the manuscript together.

**Additional information**

Supplementary information is available in the online version of the paper.

**Competing financial interests**

The authors declare no competing financial interests.

# Supplementary Information

# Antiferromagnetic ordering in van der Waals two-dimensional magnetic material MnPS$_3$ probed by Raman spectroscopy

*Kangwon Kim,*[1] *Soo Yeon Lim,*[1] *Jungcheol Kim,*[1] *Jae-Ung Lee,*[1,2] *Sungmin Lee,*[3,4] *Pilkwang Kim,*[3,5] *Kisoo Park,*[3,4] *Suhan Son,*[3,4] *Cheol-Hwan Park,*[3,5,*] *Je-Geun Park,*[3,4,*] *and Hyeonsik Cheong*[1,*]

[1] Department of Physics, Sogang University, Seoul 04107, Korea

[2] Department of Physics, Ajou University, Suwon 16499, Korea

[3] Center for Correlated Electron Systems, Institute for Basic Science, Seoul 08826, Korea

[4] Department of Physics and Astronomy, Seoul National University, Seoul 08826, Korea

[5] Center for Theoretical Physics, Seoul National University, Seoul 08826, Korea

**Contents:**

**NOTE S1.** Raman tensor analysis

**Figure S1.** Polarization dependence of peak intensities at 10 K.

**Figure S2.** Circularly polarized Raman spectra of bulk MnPS$_3$ at 290 and 10 K.

**Figure S3.** Raman spectra of bulk FePS$_3$, NiPS$_3$, and MnPS$_3$ at 290 and 10 K.

**Table S1.** Comparison of observed Raman modes with calculated Raman active modes of bulk MnPS$_3$ with antiferromagnetic ordering.

**Figure S4-5.** Calculated Raman active modes.

**Figure S6.** Polarization dependence of P$_5$ from bulk MnPS$_3$.

**Figure S7.** Temperature dependent Raman spectrum of bulk MnPS$_3$.

**NOTE S1.** Raman tensor analysis

Bulk MnPS$_3$ has a monoclinic structure with the point group $C_{2h}$ (C2/m). The irreducible representations at the $\Gamma$-point are $\Gamma = 16A_g + 14A_u + 12B_g + 18B_u$, where $A_g$ and $B_g$ modes are Raman-active. The Raman tensors for the $A_g$ and $B_g$ modes can be expressed as

$$R(A_g) = \begin{bmatrix} a & \cdot & d \\ \cdot & b & \cdot \\ d & \cdot & c \end{bmatrix}, \quad R(B_g) = \begin{bmatrix} \cdot & e & \cdot \\ e & \cdot & f \\ \cdot & f & \cdot \end{bmatrix}.$$

Linearly polarized light propagating in the $z$ direction can be represented by $e_i = \begin{pmatrix} \cos\theta \\ \sin\theta \\ 0 \end{pmatrix}$, where $\theta$ is the polarization angle. The Raman cross section is proportional to $|\langle e_s | R | e_i \rangle|^2$, where $e_i$ and $e_s$ are the polarizations of the incident and the scattered light, respectively. For the parallel polarization configuration, the intensities of $A_g$ and $B_g$ modes are proportional to

$$|\langle e_s | R(A_g) | e_i \rangle|^2 = \left| \frac{a+b}{2} + \frac{a-b}{2}\cos 2\theta \right|^2 \quad \text{and} \quad |\langle e_s | R(B_g) | e_i \rangle|^2 = |e \sin 2\theta|^2,$$

respectively. By measuring the polarization dependence of the Raman spectrum from bulk MnPS$_3$ shown in Figure S6 and fitting the intensities as a function of the polarization angle, we obtain $a \cong b$ for P$_4$ and P$_6$ and $a \cong -b$ for P$_2$, P$_3$, and P$_5$.

For circularly polarized light propagating in the z direction $\sigma\pm = \frac{1}{\sqrt{2}} \begin{pmatrix} 1 \\ \mp i \\ 0 \end{pmatrix}$, the intensity of the $A_g$ mode proportional to $|\langle \sigma + | R(A_g) | \sigma + \rangle|^2 = \frac{1}{4}|a+b|^2$ or $|\langle \sigma - | R(A_g) | \sigma + \rangle|^2 = \frac{1}{4}|a-b|^2$. On the other hand, for the $B_g$ mode, $|\langle\sigma+|R(B_g)|\sigma+\rangle|^2 = 0$ or $|\langle\sigma+|R(B_g)|\sigma+\rangle|^2 = e^2$. By using the values from linearly polarization dependence, $|\langle\sigma-|R(A_g)|\sigma+\rangle|^2 = \frac{1}{4}|a-b|^2 = 0$ for P$_4$ and P$_6$, but $|\langle\sigma+|R(A_g)|\sigma+\rangle|^2 = \frac{1}{4}|a+b|^2 = 0$ for P$_2$, P$_3$, and P$_5$. The measured circularly polarized Raman spectrum of bulk MnPS$_3$ in figure S2 is consistent with this Raman tensor analysis. P$_2$, P$_3$, and P$_5$ in the circularly polarized Raman spectrum are stronger than those in the linearly polarized Raman spectrum because both the $A_g$ and $B_g$ modes contribute to the peak intensities.

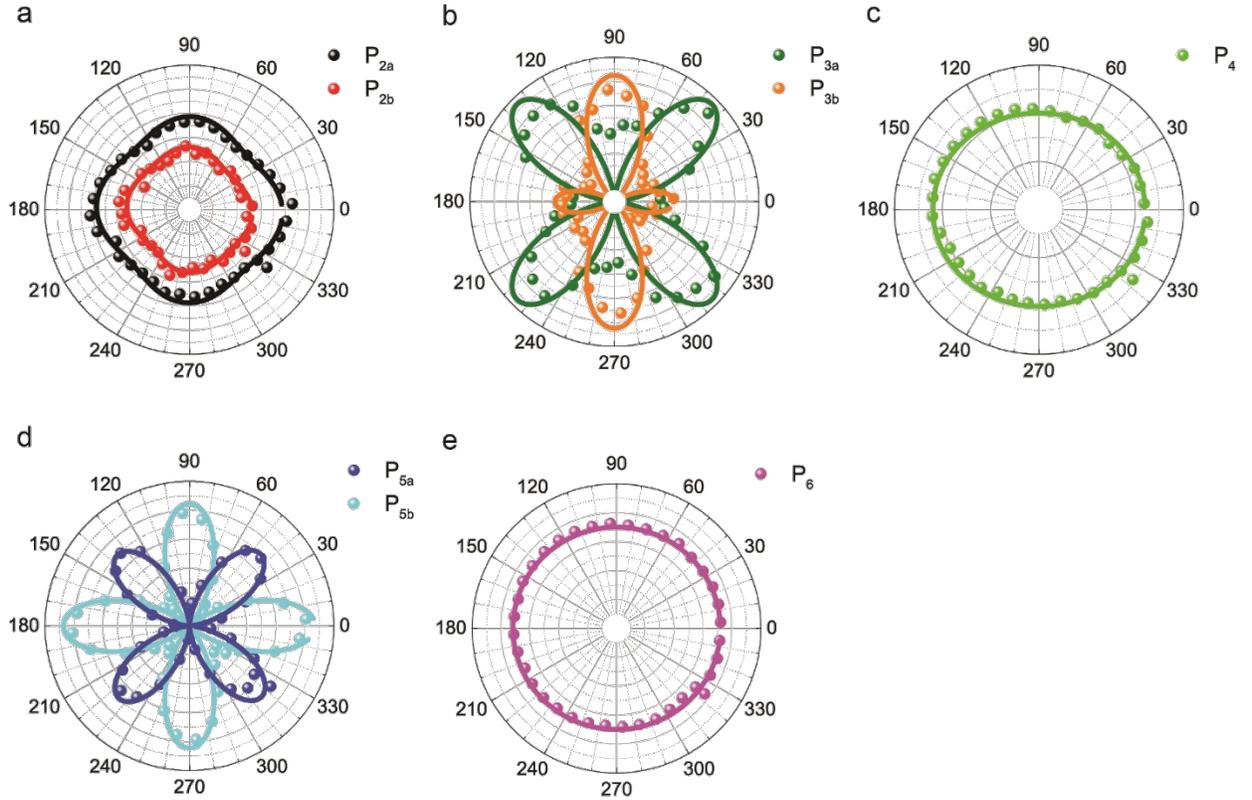

**Figure S1.** Polarization dependence of peak intensities of (a) P$_2$, (b) P$_3$, (c) P$_4$, (d) P$_5$, and (e) P$_6$ at T=10 K measured in the parallel polarization configuration. Zero degree corresponds to the $b$-axis direction of MnPS$_3$.

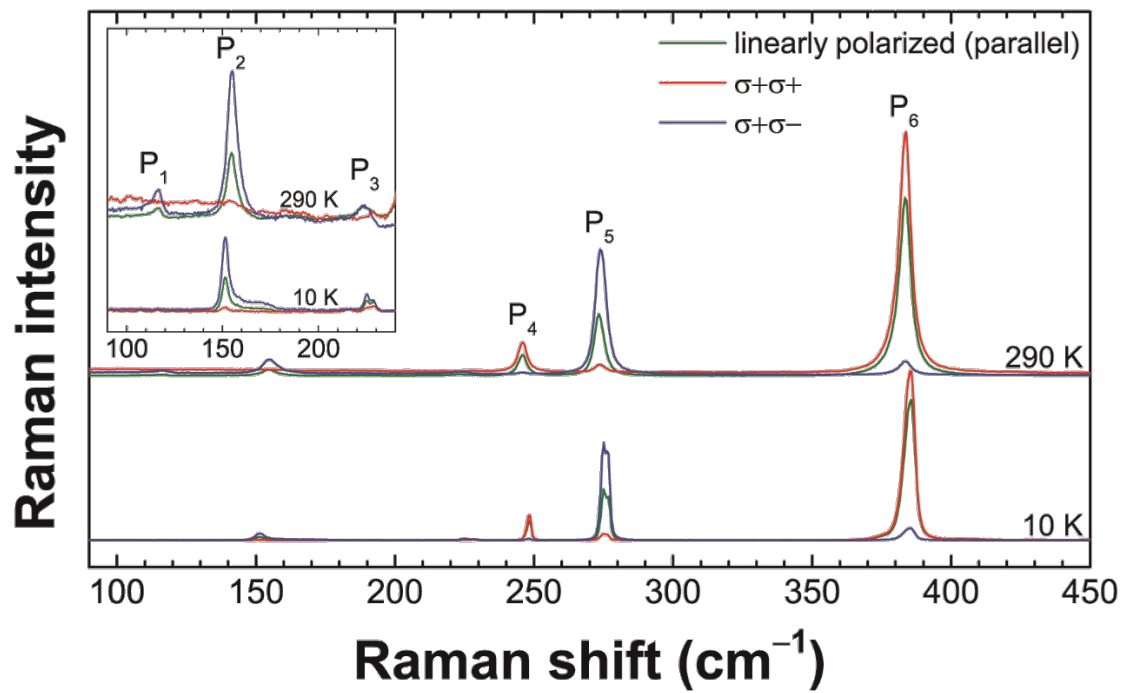

**Figure S2.** Circularly polarized Raman spectra of bulk $MnPS_3$ at 290 and 10 K.

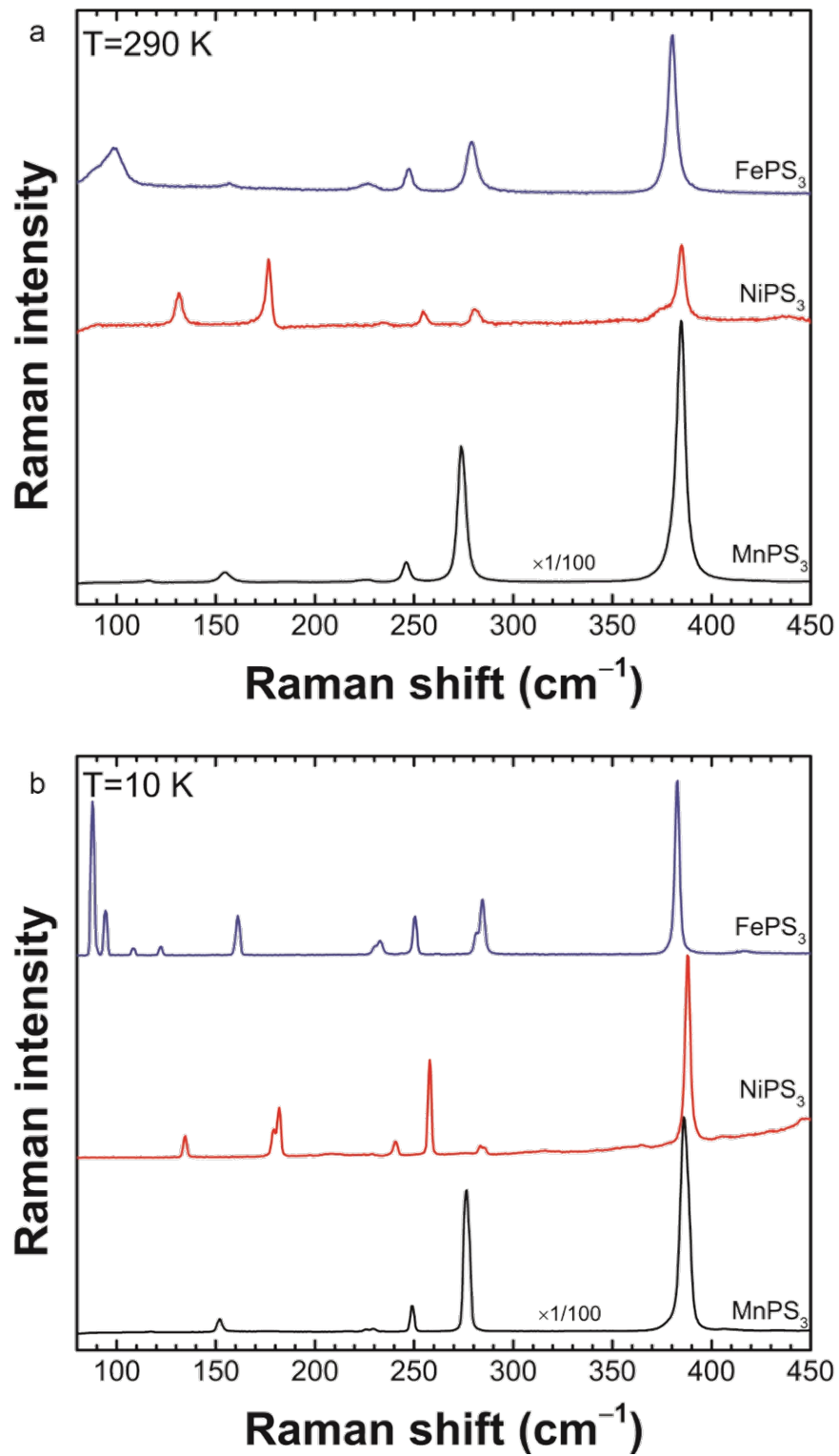

**Figure S3.** Raman spectra of bulk FePS$_3$, NiPS$_3$, and MnPS$_3$ at (a) 290 K and (b) 10 K.

**Table S1.** Comparison of observed Raman modes with calculated Raman active modes of bulk MnPS$_3$ with antiferromagnetic ordering.

| Peak | Experimental (cm$^{-1}$) | Calculation (cm$^{-1}$) | Calculated parallel intensity (a. u.) | Calculated cross intensity (a. u.) | Mode ($C_{2h}$) |
|---|---|---|---|---|---|
| P$_1$ | 115.6 (290 K) | 122.3/123.8 | 0.0 / 0.545 | 0.57 / 0.0 | B$_g$/A$_g$ |
| P$_2$ | 154.8 (290 K) 150.5 (10 K) | 160.7/161.0 | 14.06 / 0.245 | 0.16 / 13.61 | A$_g$/B$_g$ |
| P$_3$ | 223.8/227.4 (10 K) | 224.1/227.0 | 0.005 / 4.455 | 5.66 / 0.0 | B$_g$/A$_g$ |
| P$_4$ | 245.8 (290 K) 247.2 (10 K) | 262.1 | 128.42 | 0.0 | A$_g$ |
| P$_5$ | 272.9/ 274.6 (290 K) 273.7/ 275.3 (10 K) | 284.0/285.1 | 0.005 / 328.69 | 323.80 / 0.0 | B$_g$/A$_g$ |
| P$_6$ | 383.4 (290 K) 383.9 (10 K) | 385.4 | 995.41 | 0.0 | A$_g$ |

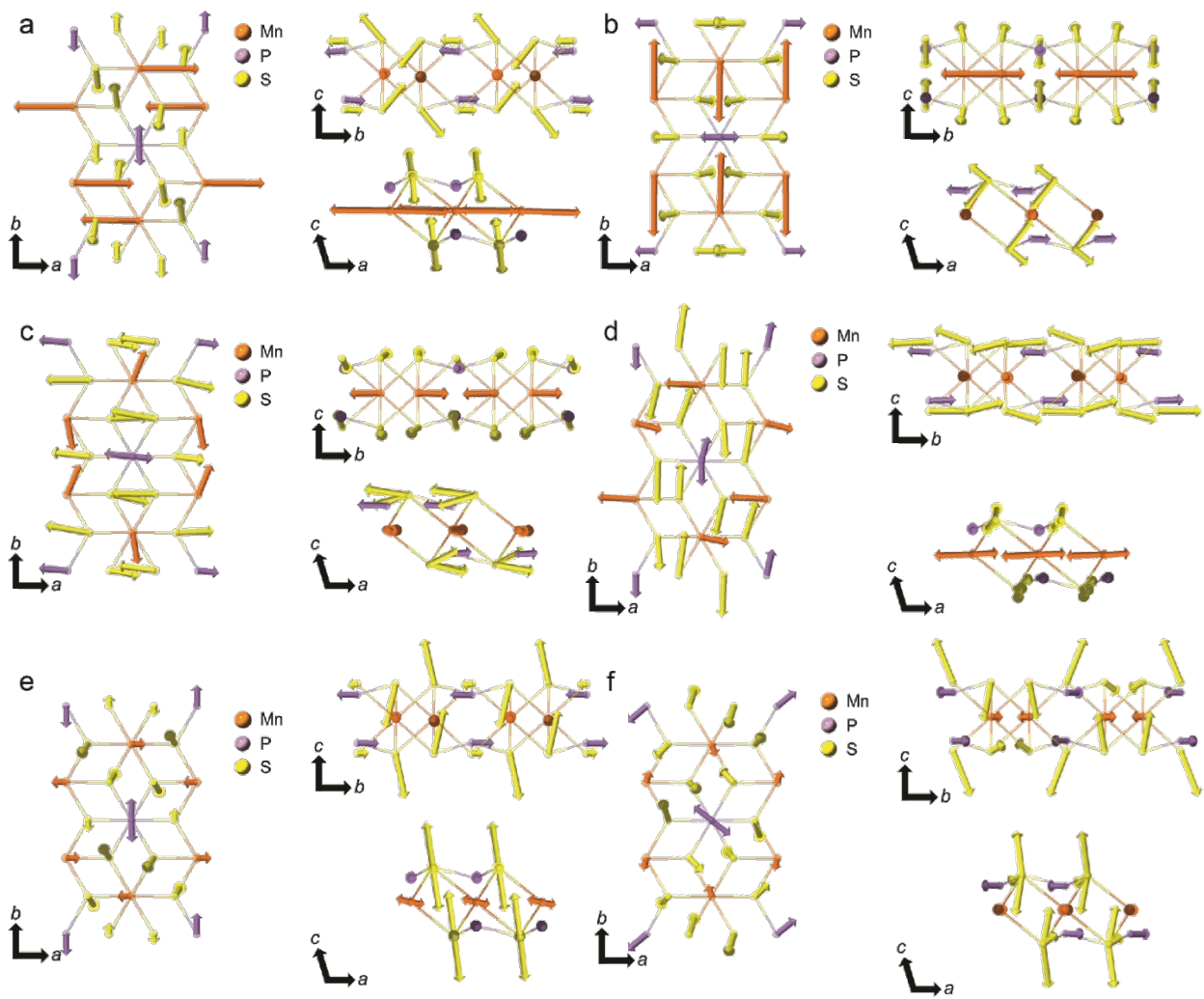

**Figure S4.** Calculated Raman active modes of $P_1$, $P_2$, and $P_3$. (a) $B_g$ and (b) $A_g$ modes for $P_1$. (c) $A_g$ and (d) $B_g$ modes for $P_2$. (e) $B_g$ and (f) $A_g$ modes for $P_3$.

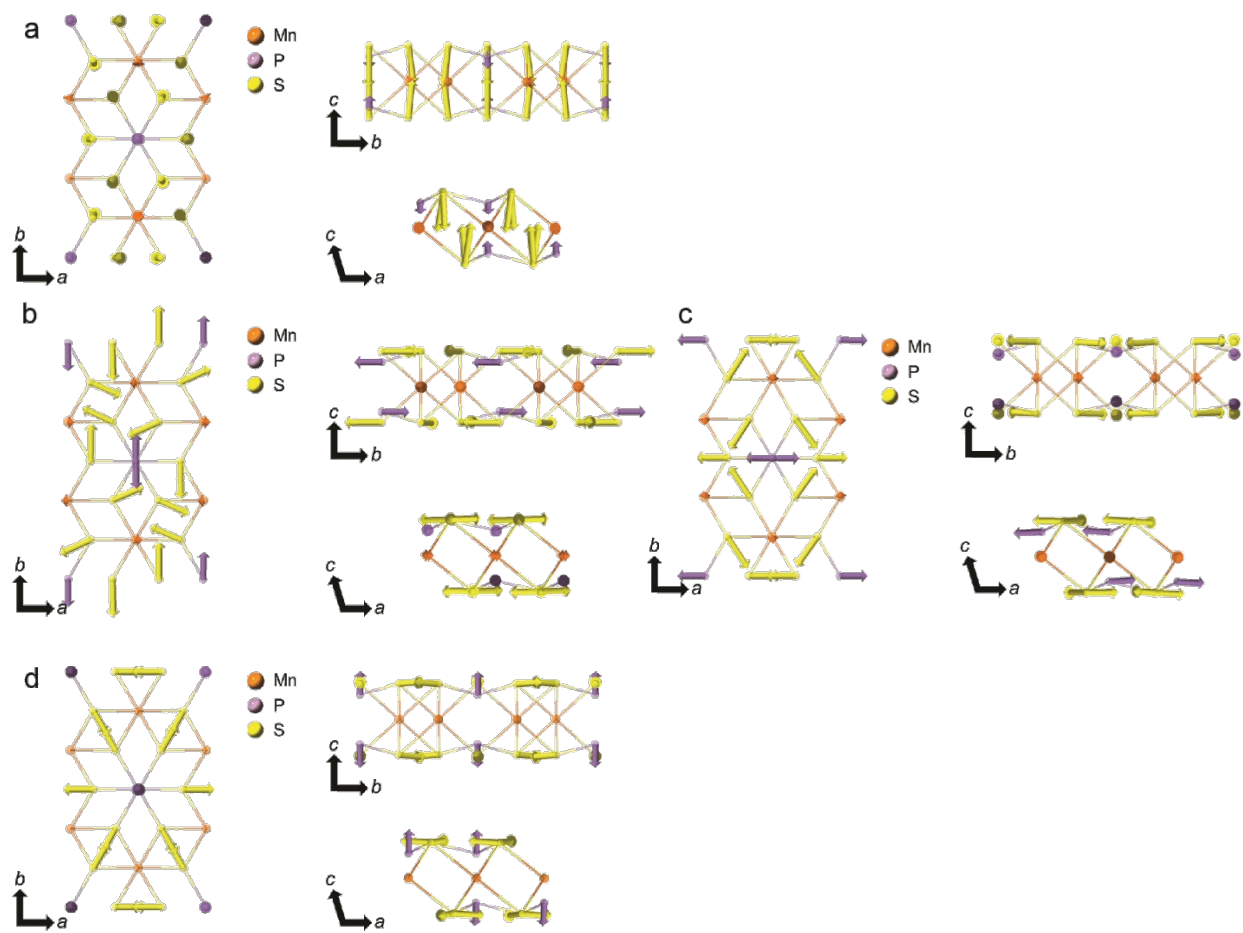

**Figure S5.** Calculated Raman active modes of $P_4$, $P_5$, and $P_6$. (a) $A_g$ mode for $P_4$. (b) $B_g$ and (c) $A_g$ modes for $P_5$. (e) $A_g$ mode for $P_6$.

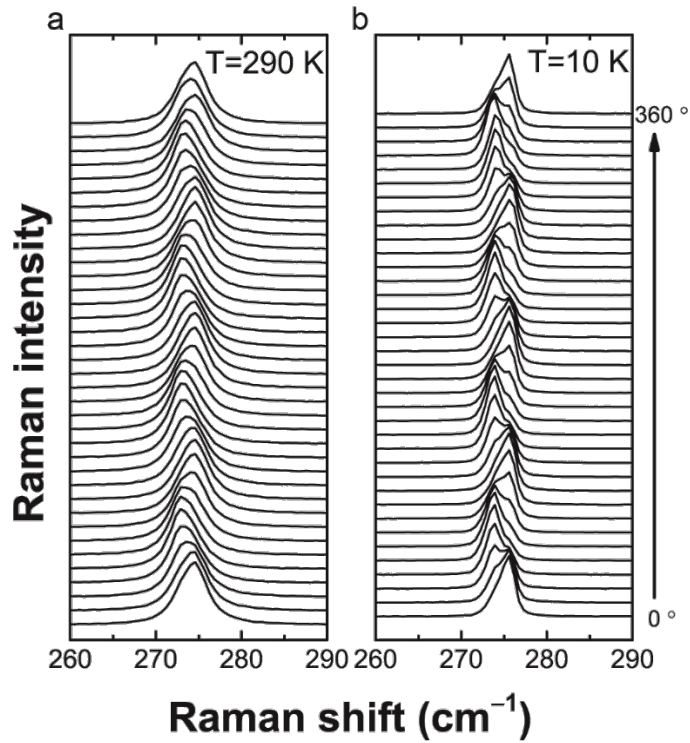

**Figure S6.** Polarization dependence of $P_5$ from bulk $MnPS_3$ at (a) 290 and (b) 10 K measured in the parallel polarization configuration (excitation and scattered photon polarizations in the same direction).

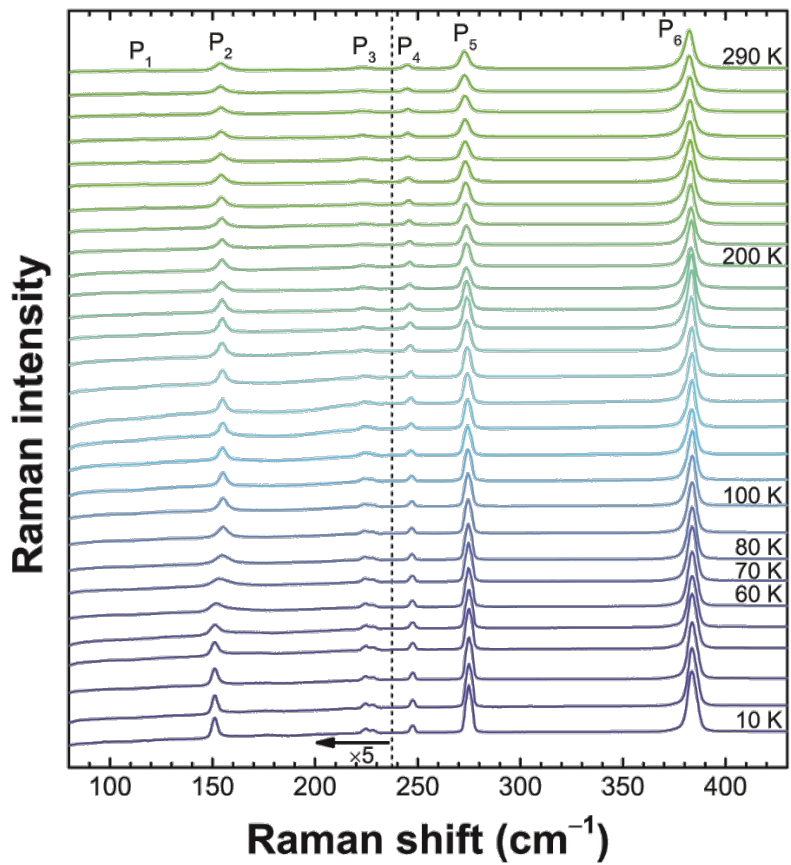

**Figure S7.** (a) Temperature dependence of the Raman spectrum of bulk MnPS$_3$ in 10 K steps.

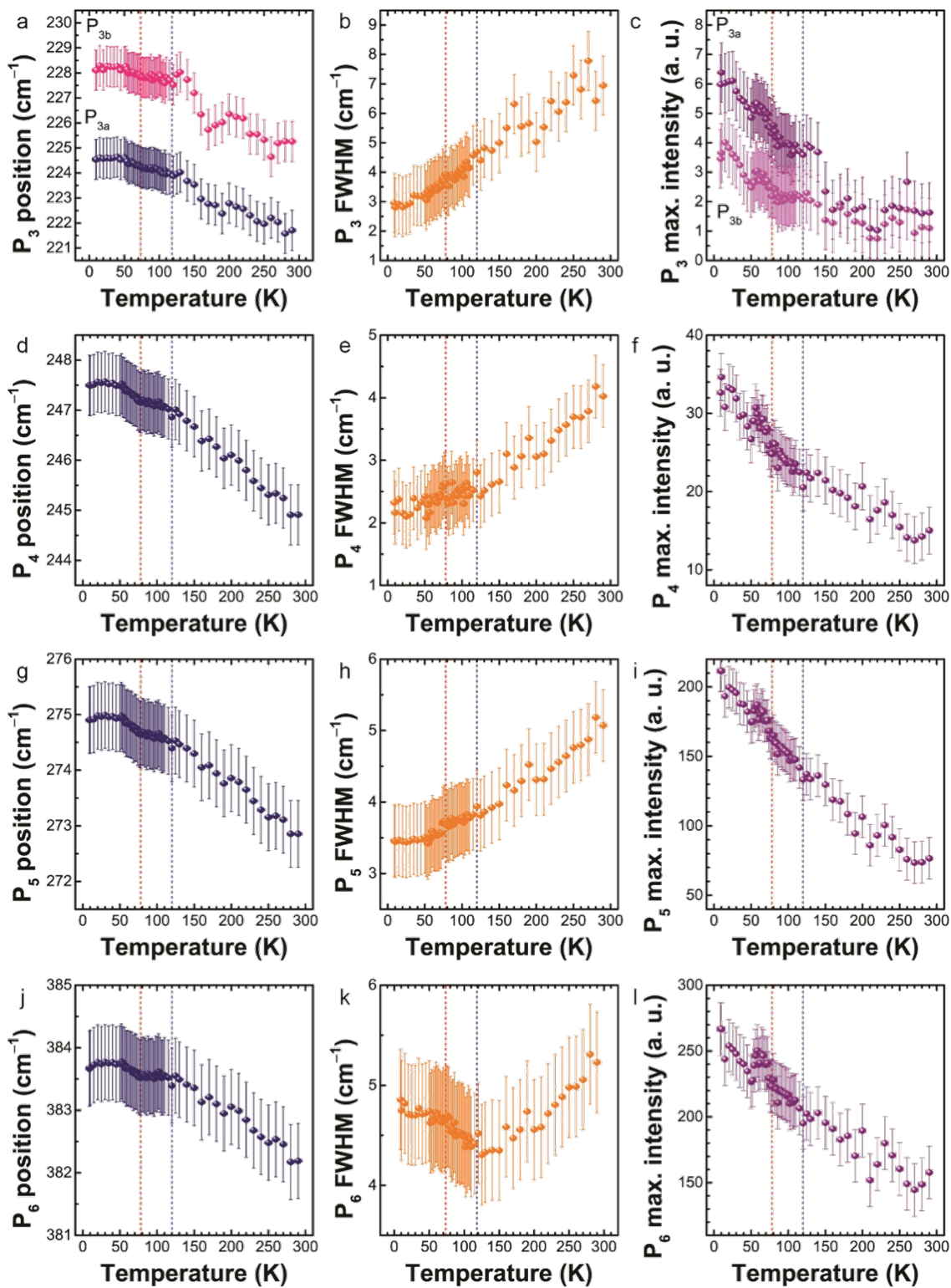

**Figure S8.** Temperature dependence of peak position, FWHM (full width at half maximum), and maximum intensity of $P_3$, $P_4$, $P_5$, and $P_6$.

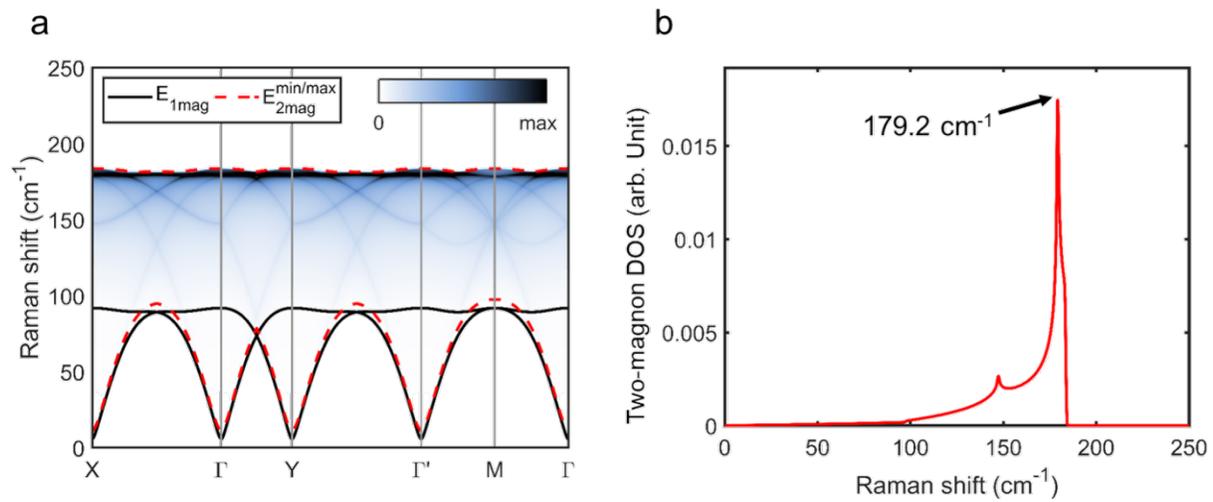

**Figure S9.** (a) Spin wave dispersion of $MnPS_3$. (b) Calculated two-magnon density of states (DOS) of $MnPS_3$.

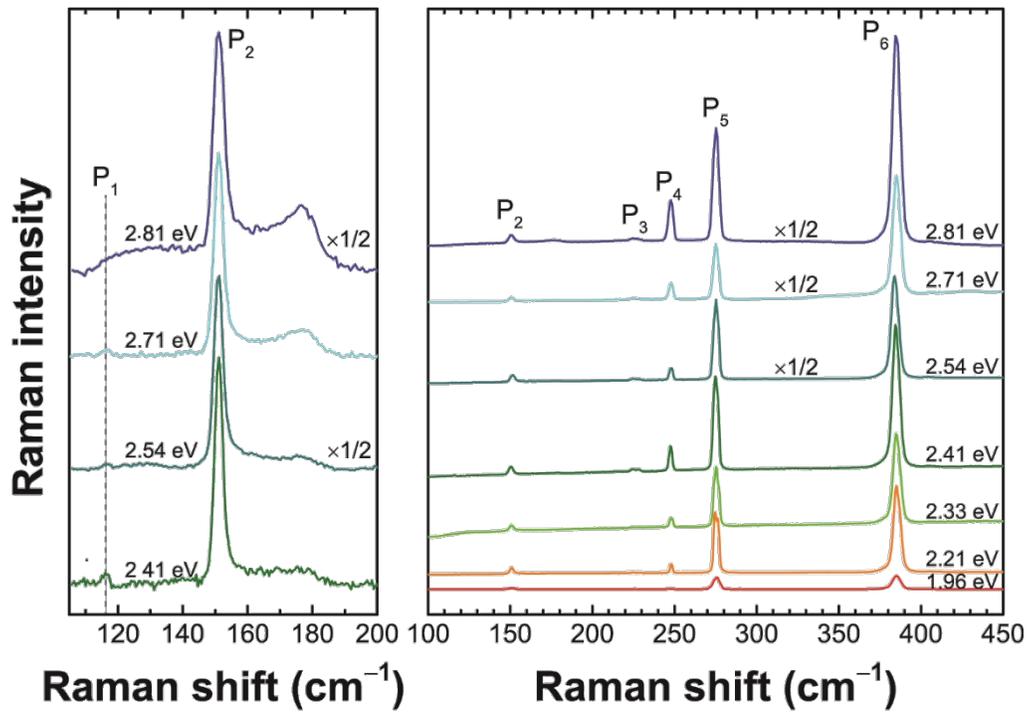

**Figure S10.** Excitation energy dependence of the Raman spectrum of bulk MnPS$_3$ at 10 K.

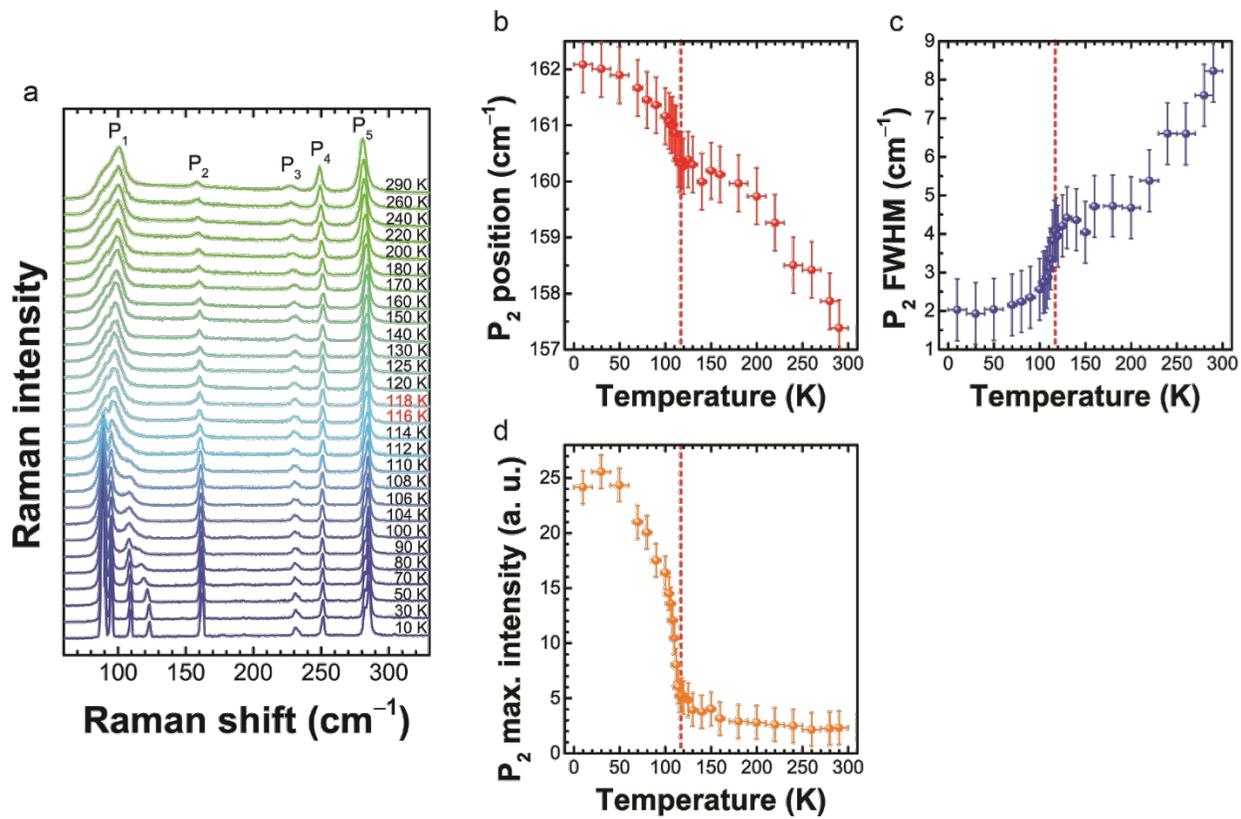

**Figure S11.** (a) Temperature dependence of the Raman spectrum of bulk FePS$_3$ from ref. [1]. (b) Peak position, (c) FWHM, and (d) maximum peak intensity of P$_2$ as a function of temperature.

**NOTE S2.** Discussion of temperature dependence for $P_1$ in bulk $MnPS_3$

In bulk $MnPS_3$, $P_1$ is already very weak at room temperature, and its intensity decreases monotonically with decreasing temperature, becoming almost unresolvable below ~140 K as shown in figure S12 when the 2.54-eV excitation is used. With the 2.41-eV excitation energy, the $P_1$ peak can be resolved at low temperatures and its temperature dependence can be analyzed. Its position and the linewidth also show abrupt change near the Néel temperature as in the case of $P_2$ (see figure S13), although the error bars are relatively large due to the very weak signal intensity. Because the $P_1$ mode also involves the vibrations of Mn ions, its coupling with the magnetic ordering is not surprising, especially since the corresponding mode was used to detect antiferromagnetic ordering in $FePS_3$ [1].

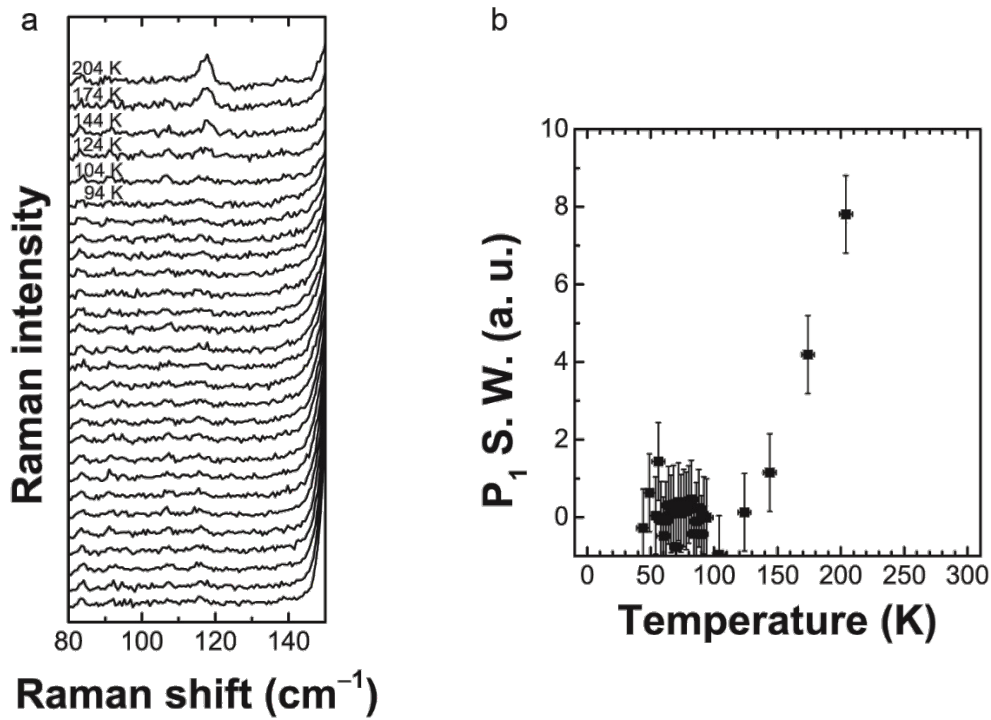

**Figure S12.** (a) Temperature dependence of $P_1$. (b) Spectral weight of $P_1$ as a function of temperature (measured with the 2.54-eV excitation).

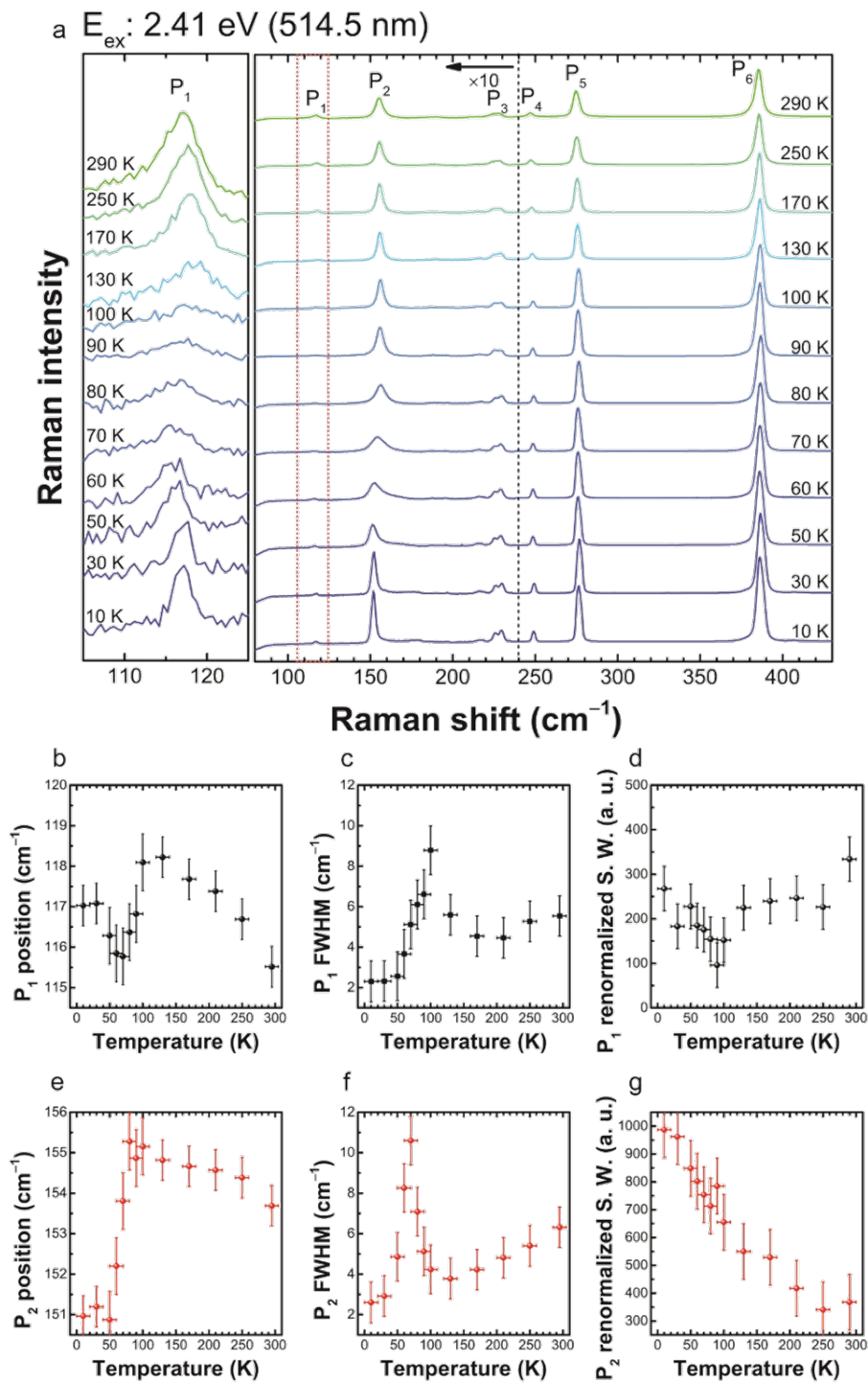

**Figure S13.** (a) Temperature dependence of the Raman spectrum of bulk $MnPS_3$ measured with the 2.41-eV excitation energy. (b,e) Peak position, (c,f) FWHM, and (d,g) renormalized spectral weight of $P_1$ and $P_2$, respectively, as a function of temperature.

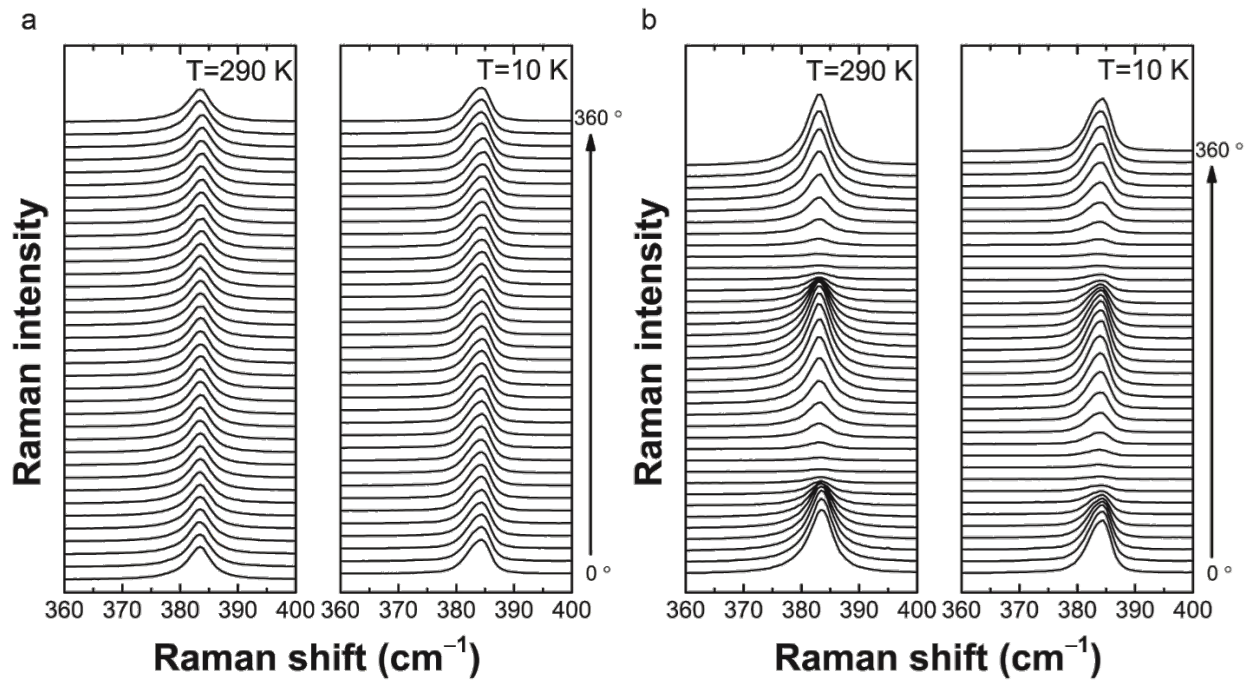

**Figure S14.** (a) Polarization dependence of $P_6$ at 290 and 10 K measured in the parallel polarization configuration. (b) Dependence of $P_6$ on the relative angle between the incident and scattered polarizations measured at 290 and 10 K.

**NOTE S3.** Monolayer MnPS$_3$

We prepared a monolayer MnPS$_3$ sample and measured the Raman spectrum at 290 K. Because the signal was very weak, we first tried to find the maximum excitation power that would not cause damage to the sample. We found that the sample starts to show damages when a 150 µW laser power was used with the sample in vacuum [figure S15(b)]. We used 100 µW of the excitation laser power for the measurements of the monolayer sample in vacuum. A monolayer MnPS$_3$ sample was identified by measuring atomic force microscope and polarized Raman spectra as shown in figure S15 (c) and (d). Monolayer MnPS$_3$ has a trigonal structure with the point group $D_{3d}$, whereas thicker MnPS$_3$ has a monoclinic structure with the point group $C_{2h}$. Both P$_3$ and P$_5$ of monolayer MnPS$_3$ comprise degenerate E$_g$ modes, whereas the degeneracy is broken in the multilayer case. Therefore, P$_3$ and P$_5$ modes are isotropic in monolayer but exhibit polarization dependence in multilayers. Due to the small splitting, the peaks appear to move as a function of the polarization. One can exploit this effect to identify a monolayer transition metal phosphorous trisulfide (TMPS$_3$) sample [1]. Figure S15(d) shows that P$_5$ indeed shows a small polarization-dependent shift for 2L and 8-nm samples, but there is no shift for the monolayer sample, confirming that indeed we have a monolayer sample. Figure S15(e) compares the Raman spectrum of monolayer MnPS$_3$ with those of other thicknesses. For this measurements we removed the analyzer for stronger signals (partially parallel-polarized configuration). However, the expanded plot in figure S15(e) shows that P$_2$ cannot be resolved in monolayer even with all the optimizations.

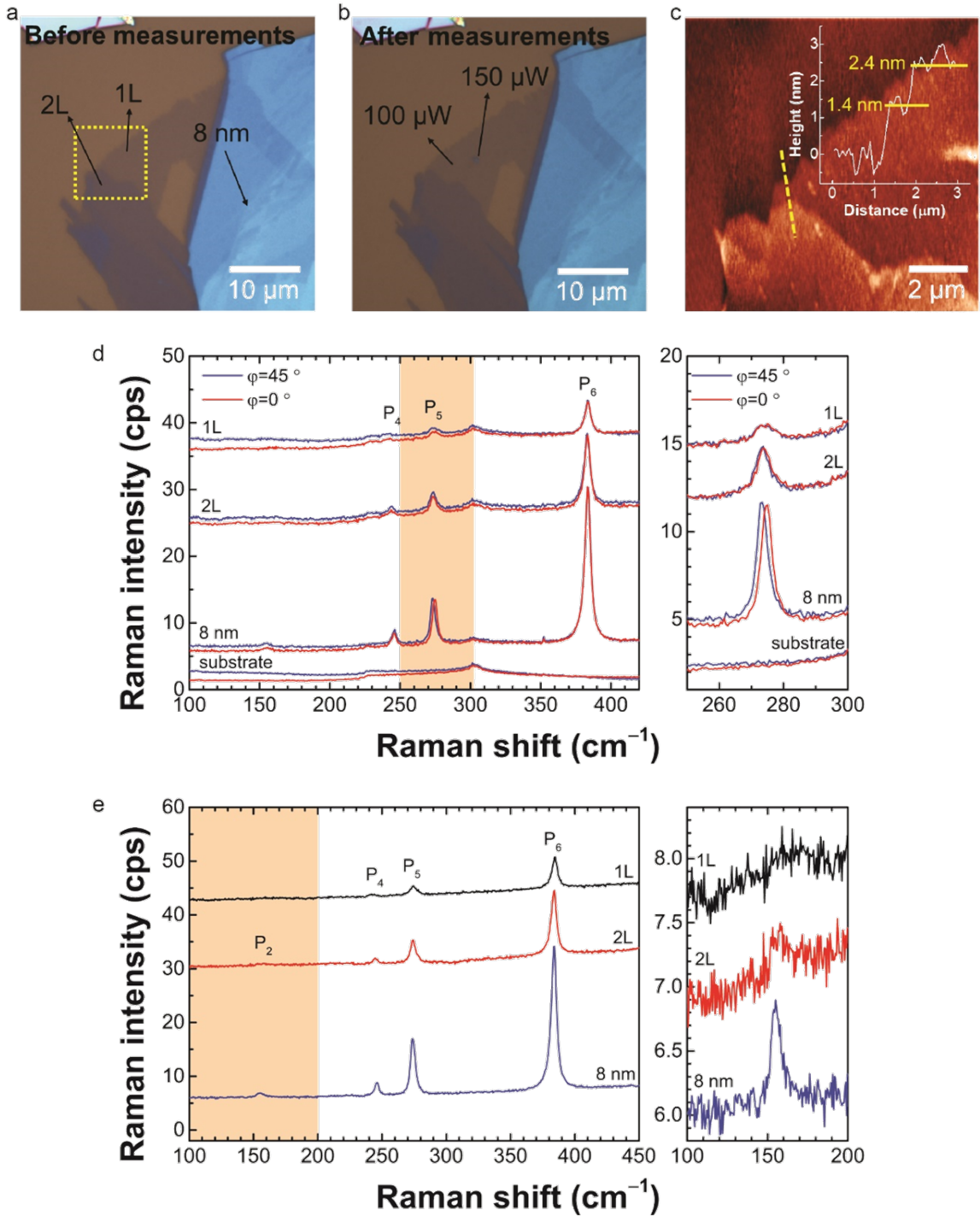

**Figure S15.** Optical contrast (a) before and (b) after measurements and atomic force microscope images of monolayer $MnPS_3$ on $SiO_2$/Si substrate. (d) Polarized Raman spectra of few-layer $MnPS_3$. φ is an angle between light polarization and *b*-axis direction of $MnPS_3$. (e) Substrate-signal-subtracted Raman spectra of few-layer $MnPS_3$.